\documentclass[aps,prb,floatfix,reprint]{revtex4-1}
\usepackage{graphicx}
\usepackage{bm,amsmath,amssymb,mathrsfs,dcolumn}
\usepackage[usenames]{color}
\usepackage{subfigure}
\usepackage{ulem}

\usepackage{color}

\usepackage{hyperref}
\hypersetup{colorlinks=true, linkcolor=blue, citecolor=red, urlcolor=blue}

\begin{document}
\title{Thermodynamic Entanglement of Magnonic Condensates}
\author{H. Y. Yuan}
\email[Electronic address: ]{yuanhy@sustc.edu.cn}
\affiliation{Department of Physics, South University of Science and Technology of China,
Shenzhen, 518055, Guangdong, China}

\author{Man-Hong Yung}
\email[Electronic address: ]{yung@sustc.edu.cn}
\affiliation{Institute for Quantum Science and Engineering and Department of Physics,
South University of Science and Technology of China, Shenzhen, 518055, China}
\affiliation{Shenzhen Key Laboratory of Quantum Science and Engineering, Shenzhen, 518055, China}

\date{\today}

\begin{abstract}
Over the last decade, significant progresses have been achieved to create 
Bose-Einstein condensates (BEC) of magnetic excitations, i.e., magnons, 
at the room temperature, which is a novel quantum many-body system with 
a strong spin-spin correlation, and contains potential applications in 
magnonic spintronics. For quantum information science, the magnonic 
condensates can become an attractive source of quantum entanglement, 
which plays a central role in most of the quantum information processing 
tasks. Here we theoretically study the entanglement properties of a magnon 
gas above and below the condensation temperature. We show that the 
thermodynamic entanglement of the magnons is a manifestation of the 
off-diagonal long-range order; the entanglement of the condensate does 
not vanish, even if the spins are separated by an infinitely large distance, 
which is fundamentally distinct from the normal magnetic ordering below 
the Curie temperature. In addition, the phase transition point occurs 
when the derivative of the entanglement changes abruptly. Furthermore, 
the spin-spin entanglement can be experimentally accessed with the 
current technology. These results provide a theoretical foundation 
for a future experimental investigation of the magnon BEC in terms 
of quantum entanglement.
%
%
%
\end{abstract}

\maketitle

{\it\color{red}Introduction.---} Magnons are the quanta of elementary
excitations in magnetically-ordered systems and behave like bosonic
quasi-particles. Information can be encoded in the excited
magnons and be transported through the magnonic spin current. This
emerging field, known as magnonic spintronics or magnoincs
\cite{Kruglyak2010,Chumak2015,Serga2010}, finds attractive applications
in information processing, due to the zero ohmic loss of magnon current and
long coherence time of magnons in low damping magnetic insulators
\cite{Kajiwara2010}. The traditional proposals in magnonics mainly
benefit from the wave nature of magnons \cite{Owens1985, Adam1988,
Lan2015, Xiansi2015, Chumak2015}, such as microwave filters, delay
lines and phase conjugators. However, the quantum nature of magnons
has not been fully explored. The technological challenge in creating
a magnon Bose-Einstein condensate (BEC) is that magnon number decreases
rapidly as the temperature decreases, and the magnon density in the
ground state is very small. However, several important experimental
progresses have been achieved over the past decade.

In 2000, the BEC of dilute magnons was proposed to explain the field-induced
N\'{e}el ordering in the spin-gap magnetic compound $\mathrm{TlCuCl}_3$,
where the magnon number was kept conserved temporality~\cite{Osawa1999, Nikumi2000}.
In 2006, magnon BEC was experimentally realized by parametrically pumping
magnons into a magnetic insulator yttrium iron garnet (YIG) using microwave
\cite{Demokritov2006}, where the magnon-magnon interaction relaxes much faster than the
magnon-lattice interaction; consequently, there is a transient time of magnon
number conservation, which provides a prerequisite for magnon BEC.
In 2011, a low-temperature BEC of magnons was experimentally observed in
gadolinium nanoparticles
\cite{Mathew2011}.

Besides, it has been proposed~\cite{Bender2012,Flebus2016} that magnon condensates
can be created through electronic pumping in a FM/normal metal (NM) bilayer,
where the ground state can maintain a macroscopic number of strongly-correlated
magnons. This system may serve as an excellent platform for studying the quantum
properties of the system, such as spin-spin entanglement~\cite{Bennett2014},
which motivates our study.

Entanglement is a measure of quantum correlation between two or more quantum systems,
which has attracted significant attention due to its intriguing applications in quantum
information science~\cite{Amico2008,Horodecki2009, Nielsen,Kassal2011,Yung2014,Zeng2015}.
The studies of entanglement in condensed-matter physics have become fruitful,
covering many physical aspects. For example, the scaling behavior of entanglement
in the vicinity of transition point of a magnetic system~\cite{Osterloh2002, Vidal2003},
the entanglement area law in the superfluid phase of Helium-3 ~\cite{Herdman2017,Laflorencie2016},
and quantification of entanglement in cold atom many-body systems~\cite{Cramer2011}.
However, entanglement is usually a very fragile quantum property, which would be
eliminated through decoherence i.e., the interaction with the environment~\cite{Schlosshauer2005,Leic}.

In this letter, we focus on the spin-spin entanglement in a many-body magnetic 
system around the transition temperature of the magnon BEC. We show that the spin 
particles can still be entangled in a magnon condensate even if the spins are separated 
far apart in the condensate. This property is very different from the normal thermal 
states of magnetic systems, where entanglement decays exponentially, and is a 
manifestation of the long-range order of the magnon BEC.

Moreover, our results provide a direct method to access the spin entanglement 
through the measurement of magnon density. Previously, the spatial entanglement 
of a normal (non-magnetic) BEC was studied~\cite{Heaney2007}; there it was 
proposed that the entanglement can be extracted with interacting with an 
external probe, which is inefficient. Here the magnetic degrees of freedom 
in the magnon BEC can be directly probed, which can be achieved with the 
current technology. For example, one can measure the magnon density in a 
pumped magnetic system~\cite{Demokritov2006}, or measure the spin-spin 
entanglement directly in ultracold systems in an optical lattice ~\cite{Fukuhara2015}.


Let us start with a group of interacting spins described by the following Hamiltonian:
\begin{equation}
\mathcal{H}=-J\sum_{\langle ij \rangle} \mathbf{S}_i \cdot \mathbf{S}_j
- \sum_{i} \mathbf{S}_{i} \cdot \mathbf{B} + H_{\mathrm{ani}},
\label{hamiltonian}
\end{equation}
where $\mathbf{S}_i$ is the spin operator on the $i$-th site, $J$ is
exchange constant, $\mathbf{B}$ is an external field. The first term captures
the exchange energy and the sum is taken for nearest neighbours; the second
term represents the Zeeman energy, and the third term $H_{\mathrm{ani}}$ denotes the
anisotropy of the system. The ground state of Hamiltonian (\ref{hamiltonian})
is a ferromagnet domain where all the spins align along the direction of magnetic
field.

To take a step further, let us assume that the magnon excitations are dilute,
which means that the low-energy excitation can be described by the bosonic
Hamiltonian: $\mathcal{H}=\sum_q \hbar \omega_q a^\dagger_q a_q$,
where $\hbar$ is Planck constant, $\omega_q$ is the magnon frequency,
$a^\dagger_q$, $a_q$ are the creation and annihilation operators for magnons
with wave vector $q$; they obey the bosonic commutation relations $[a_q, a^\dagger_{q'}]=\delta_{qq'}$.

In the following, we first illustrate the theory to deal with the two-spin
entanglement in a spin system containing purely thermal magnons, then generalize
the theory to a system with both thermal and condensed magnons.

{\it\color{red}Thermal magnons.---} The quantum state of two spins at
$i$-th and $j$-th cite of the magnetic system can be described in terms
of the reduced density matrix $\rho_{ij}$ obtained by tracing over
all the other spins in the ground state. In the standard basis,
($|\uparrow \uparrow \rangle, |\uparrow \downarrow \rangle,|\downarrow
\uparrow \rangle,|\downarrow \downarrow \rangle$), the two-spin density
matrix can be written as

\begin{equation}
\rho_{ij}=
\left ( \begin{array}{cccc}
\langle \kappa^+ _i \kappa^-_j\rangle & \langle \kappa^+_i \sigma^-_j\rangle
&\langle \sigma^-_i \kappa^+_j\rangle & \langle \sigma^-_i \sigma^-_j\rangle \\
\langle \kappa^+ _i \sigma^-_j\rangle & \langle \kappa^+_i \kappa^-_j\rangle
&\langle \sigma^-_i \sigma^+_j\rangle  & \langle \sigma^-_i \kappa^+_j\rangle\\
\langle \sigma^+ _i \kappa^+_j\rangle & \langle \sigma^+_i \sigma^-_j\rangle
&\langle \kappa^-_i \kappa^+_j\rangle  & \langle \kappa^-_i \sigma^-_j \rangle\\
\langle \sigma^+_i \sigma^+_j\rangle & \langle \sigma^+_i \kappa^+_j \rangle
&\langle \kappa^-_i \sigma^+_j\rangle  & \langle \kappa^-_i \kappa^-_j\rangle\\
\end{array} \right ),
\label{rho}
\end{equation}
where $\kappa^\pm_i= \frac{1}{2}(1\pm\sigma^z_i), \sigma^\pm_i =
\frac{1}{2}(\sigma^x_i \pm i\sigma^y_i)$, $\sigma^x_i,\sigma^y_i, \sigma^z_i$
are the Pauli matrix describing the $i$-th spin. To simplify the density
matrix, we consider the ground state that all the spins align along the
direction of external field ($z-$axis) and the system has a rotational
symmetry around $z-$axis. Then the density matrix could be simplified
as~\cite{note02,Wooters1998}
\begin{equation}
\rho_{ij}=
\left ( \begin{array}{cccc}
\Lambda_{00} & 0&0 & 0 \\
0 & \Lambda_{11} &\Lambda_{12}  & 0\\
0 & \Lambda_{12} &\Lambda_{22}  & 0\\
0 & 0 &0  & \Lambda_{33}\\
\end{array} \right ),
\end{equation}
where $\Lambda_{00},\Lambda_{11},\Lambda_{12},\Lambda_{22},\Lambda_{33}$
are the corresponding matrix element in matrix $(\ref{rho})$.
The amount of entanglement between the $i-$th spin and $j-$th spin
can be quantified by calculating the Wooters' concurrence defined
as \cite{Wooters1998},
\begin{equation}
C_{ij}=\max (0,\lambda_1-\lambda_2-\lambda_3-\lambda_4),
\end{equation}
where $\lambda_1,\lambda_2,\lambda_3,\lambda_4$ are the square root
of the eigenvalues of $\rho_{ij}[ (\sigma^y \otimes \sigma^y)
\rho^*_{ij} (\sigma^y \otimes \sigma^y)]$ in a non-increasing order.
Here $\lambda_{1,2} = \sqrt{\Lambda_{11}\Lambda_{22}} \pm |\Lambda_{12}|,
\lambda_{3,4}=\sqrt{\Lambda_{00}\Lambda_{33}}$,
and the concurrence is given by \cite{note01}
\begin{equation}
\begin{aligned}
C_{ij}=\frac{1}{2} \max (0,|\langle \sigma_i^x \sigma_j^x + \sigma_i^y \sigma_j^y \rangle|\\
-\sqrt{(1+ \langle \sigma_i^z \sigma_j^z \rangle)^2-\langle \sigma_i^z + \sigma_j^z \rangle^2} ).
\end{aligned}
\end{equation}

Now, applying the Holstein-Primakoff transformation (HPT) for converting 
spin operators to bosonic operators~\cite{HP1940}, 
$S^+_i= \sqrt{2S-a^\dagger_ia_i} \ a_i, S^-_i= a^\dagger_i \sqrt{2S-a^\dagger_ia_i},
S^z_i = S-a^\dagger_i a_i$, where $S^\pm_i = \sigma_i^\pm /2, S^z_i = \sigma_i^z/2$,
$a_i, a^\dagger_i$ are the magnon creation and annihilation operators
in the real space that obey bosonic commutation relations, the concurrence
can be reduced to the one-particle reduced density matrix
(1-RDM) of the system \cite{Yang1962,Lieb2000} i.e.
$C_{ij}=\langle a_i^\dagger a_j \rangle$. Here we have assumed that
the average magnon density is very small ($\langle a^\dagger_i a_i \rangle \ll 2S$)
such that it is reasonable to expand the HPT to the linear order of
magnon creation (annihilation) operator. Since this approximation is
valid for a magnetic system well below the Curie temperature ($T_c$),
our results presented below is applicable at this regime. However, for
systems with a strong anisotropy and/or strong applied fields, the spin
wave gap becomes sufficiently large, limiting the number of thermal
magnons; our results can become applicable at higher temperatures.

Through a Fourier transform of the magnon operators, i.e.,
$a_i = 1/\sqrt{N_s}\sum_q e^{-iq\cdot R_i}a_q$,
the concurrence is recasted in the following form:
\begin{equation}
\begin{aligned}
C_{ij}=\frac{4S}{N_s} \left|\sum_q e^{iq\cdot R_{ij}}n_q \right |,
\end{aligned}
\label{discrete}
\end{equation}
where $N_s$ is the total number of spins in the system,
$n_q = \langle a^\dagger_q a_q \rangle$ is the density of magnons with
wave vector $q$, $R_{ij}=R_j-R_i$ is the relative distance between
$i-$th spin and $j-$th spin. Translational symmetry of the system is
used in the derivation such that the concurrence only depends on the
relative distance between the two spins. If only one-magnon excitation
is considered, $n_q =1$, then the sum gives zero entanglement of two
different spins in the thermal dynamic limit, which is qualitatively
consistent with the literature \cite{Asoudeh2004}. Generally, the magnon
distribution obeys the Bose-Einstein statistics,
$n_q = 1/ (e ^{\hbar \omega_q/ k_BT} -1)$, where the dispersion relation
$\omega_q/\gamma=H_{\mathrm{ex}} q^2 + B + H_{\mathrm{an}}$,
$H_{\mathrm{ex}}$ is exchange field, $H_{\mathrm{an}}$ is anisotropy field,
$\gamma$ is gyromagnetic ratio and $k_B$ is Boltzmann constant.

In the continuum limit, the sum in Eq. (\ref{discrete}) could be
replaced by integral in the momentum space i.e. $\sum_q \rightarrow V \int 4\pi q^2 dq$.
Given that $n_q$ is an even function of $q$, due to the $q^2$ terms in $\omega_q$,
the concurrence becomes
\begin{equation}
\begin{aligned}
C_{ij}=\frac{1}{\pi^2} \left| \int_0^\infty \frac{p^2 \cos r p}{z_0 e^{bp^2} - 1}dp \right |,
\end{aligned}
\label{continum}
\end{equation}
where $p=qd$ is a dimensionless quantity, $d$ is lattice constant,
$r=|R_{ij}|$, $z_0 = \exp[\hbar \gamma (B+H_{an})/ (k_BT)]$,
$b= \gamma H_{\mathrm{ex}}/ (k_B T)=T_c/T$ and $T_c = \gamma H_{\mathrm{ex}}/k_B$
is the approximated Curie temperature. For small fields and weak anisotropy,
$z_0 \approx 1$, the integral can be evaluated analytically by only
considering the excitation of long wavelength magnons such that,
$e^{bp^2} \sim b p^2 + b^2p^4/2$,
\begin{equation}
C_r\equiv C_{i,i+r} = \frac{\sqrt{2}}{2\pi} e^{-r/\xi_e}\left ( \frac{T}{T_c}\right )^{3/2}
\end{equation}
where $\xi_e =d\sqrt{T_c/2T}$ is defined as entanglement length of
the system.

\begin{figure}[t]
\includegraphics[width=0.8\columnwidth]{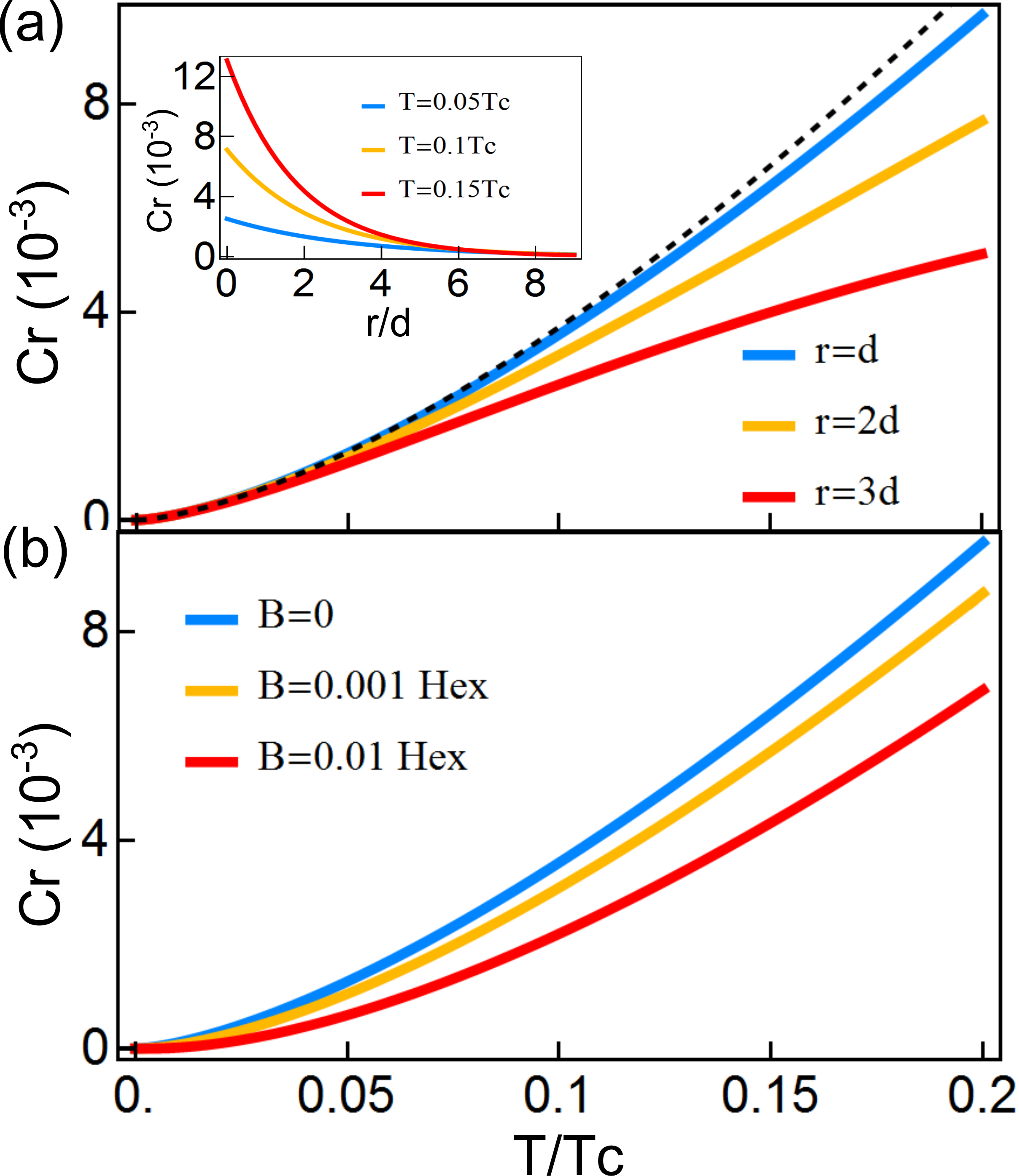}
\caption{(color online) Concurrence as a function of temperature for
$r=d$ (blue line), $2d$ (yellow line), and $3d$ (red line), respectively.
The dashed line represents the reduced number of magnons as a function of $T$.
The inset shows the concurrence as a function of the distance between the two spins
for T = $0.05\mathrm{T_c}$ (blue line), $0.10\mathrm{T_c}$ (yellow line) and
$0.15\mathrm{}\mathrm{T_c}$ (red line), respectively.
(b) Concurrence of nearest spins as a function of temperature for external
fields $B=0$ (blue line), 0.001 $H_{\mathrm{ex}}$ (yellow line) and
0.01 $H_{\mathrm{ex}}$ (red line), respectively.}
\label{fig1}
\end{figure}

Figure \ref{fig1} visualizes the temperature dependence
of concurrence for $r=d$, $2d$ and $3d$, respectively. Regardless
of the distance between the two spins, the concurrence is zero at $0$ K
and then increases monotonically as the temperature increases.
This is because the spins are perfectly aligned along $z-$axis and 
the bi-spin state is a separable state $|\uparrow \uparrow \rangle$
when $T=0$ K. At finite temperatures, the magnons are excited and
the weighted sum of magnon density gives the amount of two-spin
entanglement as indicated by Eq. (\ref{discrete}).
According to the Bose-Einstein statistics, the magnon number decays
exponentially with the magnon energy, then the higher-energy
terms gives a negligible contribution to the concurrence.
For low energy magnons, the cosine factor in Eq. (\ref{discrete})
is almost equal to 1, and the concurrence is equivalent to the magnon
density. The higher the temperature, the larger the density of low
energy magnons, hence the larger the two-spin entanglement.
This argument naturally gives an estimate of the concurrence
$C_{ij} \approx 4S/N \sum_q n_q$, as shown by a dashed line in
Fig. \ref{fig1}a. This line roughly captures the temperature dependence
of the entanglement between two nearest spins.

For external fields comparable with $k_B T/\gamma$, $z_0 >1$, the
temperature dependence of concurrence could be derived by calculating
the integral Eq. (\ref{continum}) numerically, as shown in Fig. \ref{fig1}b.
The general trend of the $C_r$ vs $T$ is not qualitatively altered
by the finite fields. Nevertheless, as field increases, the entanglement
of the two spins decreases which finally approaches zero for infinitely
large field. This is because the magnon gap ($\gamma B$) keeps increasing
with the field and it becomes harder to excite magnons for larger fields,
then the two-spin state approaches a separable state
$|\uparrow \uparrow \rangle$ with zero entanglement.


\begin{figure}[t]
\includegraphics[width=0.9\columnwidth]{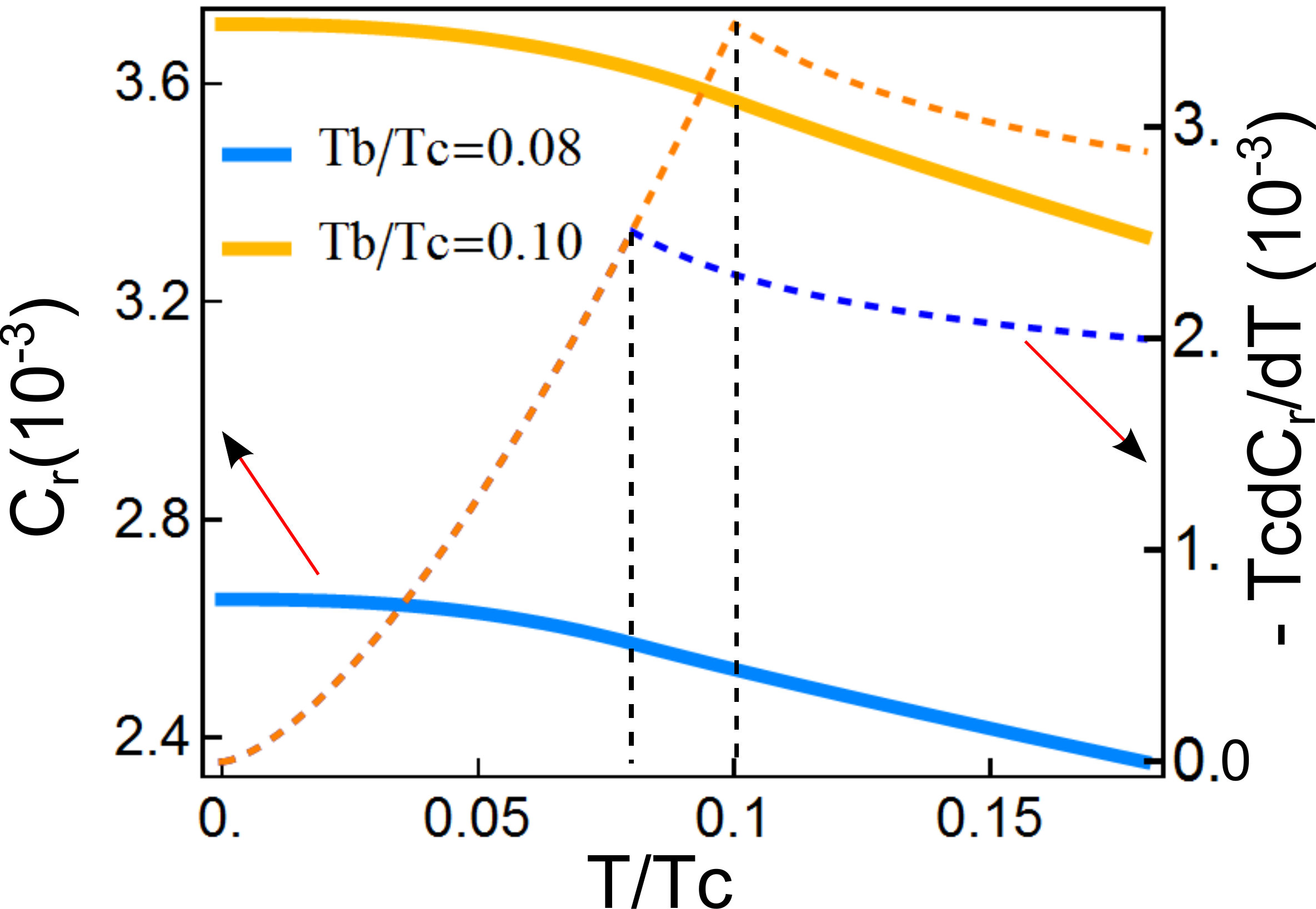}
\caption{(color online) (a) Concurrence as a function of temperature
(left axis) for $T_b/T_c =0.08$ (blue line), 0.1 (yellow line) and
the corresponding slopes of the curve are indicated as dashed lines.
The black dashed lines indicate the position of condensation temperature
($T_b$).}
\label{fig2}
\end{figure}

{\it\color{red}Condensed magnons.---}Up to this stage, our theory are
restricted to the magnons gas without magnon number conservation.
To achieved magnon BEC, it is essential to produce a magnon gas with
conserved number of magnons, at least temporarily. This may be
realized by tuning the magnetic field, lowering the temperature in some
special materials \cite{Osawa1999, Nikumi2000} and parametric pumping.
To describe such a magnon gas with fixed magnon number, the Hamiltonian
should be modified as
$\mathcal{H}=\sum_q (\hbar \omega_q -\mu) a^\dagger_q a_q
=\sum_q (\hbar \gamma H_{ex} q^2 -\mu_{\mathrm{eff}}) a^\dagger_q a_q$,
where $\mu$ is the chemical potential, and $\mu_{\mathrm{eff}}
\equiv \mu-\hbar \gamma (B+H_{an})$ is the effective potential
that subtracts the bottom of magnon band. Similar to the BEC of atoms,
the number of magnons in the spin system can be written as the sum of
magnon number in the ground state and the excited state,
i.e. \cite{Huang1963}
\begin{equation}
N=N_0 + (4\pi b)^{-3/2}N_sg_{3/2} (z),
\end{equation}
where $N_0$ the magnon number in the ground state and
$z=\exp(-\mu_{\mathrm{eff}})$ is the fugacity,
$g_{3/2} (z) = \sum_{k=1}^{\infty} z^kk^{-3/2} $
is the Bose-Einstein integral.

Above $T_b$, the magnon population in the ground state is neglectable
such that $N_0 \ll N$, then $N/N_s=(4\pi b)^{-3/2}g_{3/2} (z)$.
Below $T_b$, the effective chemical potential is zero ($z=1$) and
the occupancy of ground state is comparable with the total number of
magnons, where the condensation temperature is determined by the equality
$N/N_s=(4\pi b_{T=T_b})^{-3/2}g_{3/2} (1)$. Combining the two limits,
the chemical potential can be readily determined from a set of
self-consistent equations:
\begin{equation}
\left \{ \begin{array}{l}
\mu_{\mathrm{eff}} = k_B T \ln z \\
T=T_b  \left [g_{3/2} (1)/g_{3/2} (z)\right ]^{2/3} ,\\
\end{array} \right .
\label{consistent}
\end{equation}
where the first equation is the definition of fugacity and the second
relation is obtained from the consistent calculation of magnon density
below and above $T_b$. Once the chemical potential is determined, the
entanglement of two spins could be calculated through the integral
similar to Eq. (\ref{continum}), for $T>T_b$,
\begin{equation}
\begin{aligned}
C_{r}=\frac{1}{\pi^2} \left| \int_0^\infty \frac{p^2 \cos r p}{z^{-1} e^{bp^2}
- 1}dp \right | \ .
\label{lttb}
\end{aligned}
\end{equation}

\begin{figure}[t!]
\includegraphics[width=0.85\columnwidth]{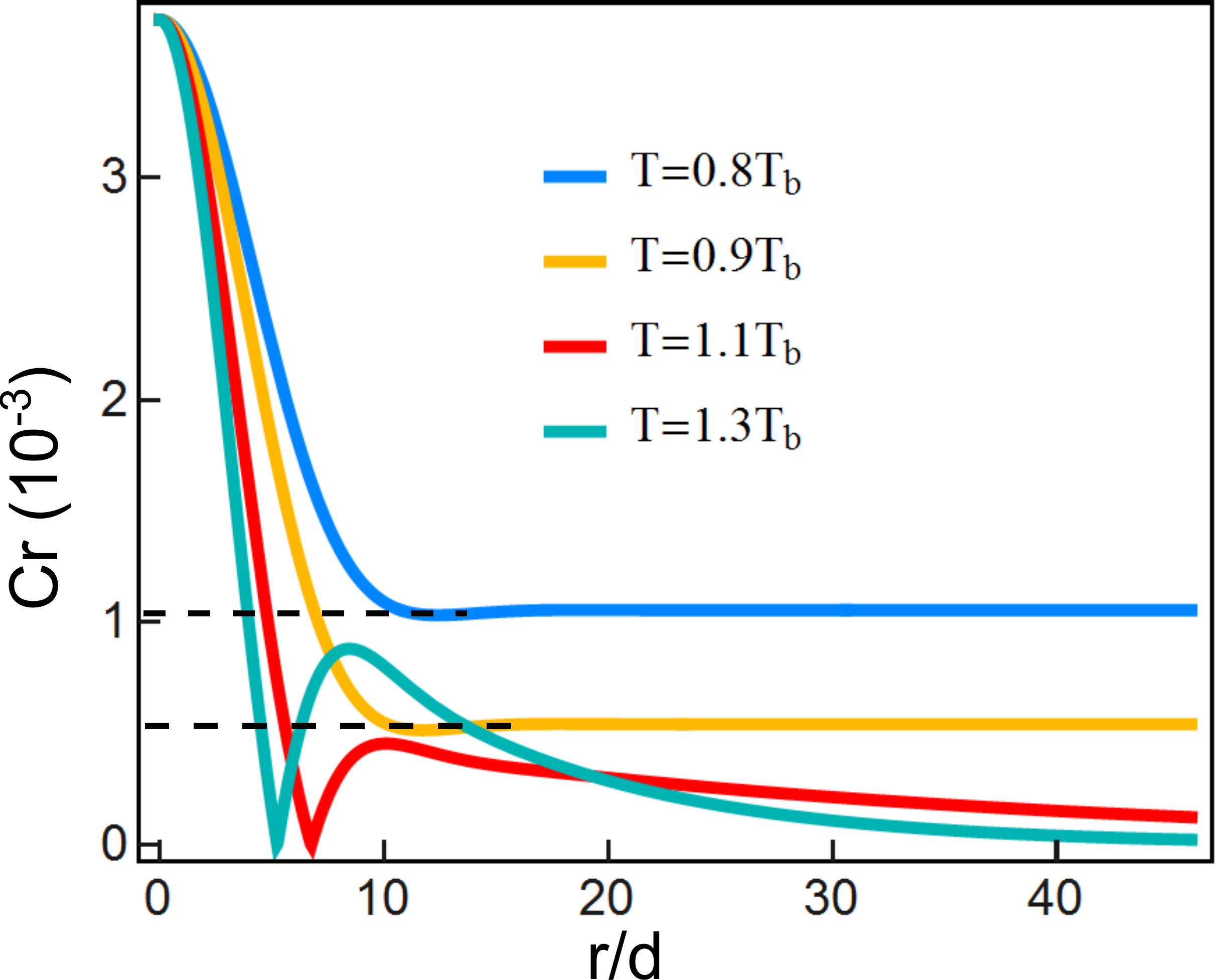}
\caption{(color online) Bi-spin entanglement as a function of
the distance between two spins at various temperature $T=0.8\mathrm{}\mathrm{T_b}$ (blue line),
$0.9\mathrm{T_b}$ (orange line), $1.1\mathrm{T_b}$ (red line), $1.3\mathrm{T_b}$ (cyln line).
 The dashed lines indicate the positions of
plateau for two well-separated spins.}
\label{fig3}
\end{figure}

Below $T_b$, this integral fails since significant number of spins
are in the ground state and they have to be considered separately
from the magnons in the excited state. By separating the magnon number
in the ground state from Eq. (\ref{lttb}), we obtain
\begin{equation}
\begin{aligned}
C_{r}=\left|\frac{2N_0}{N_s} + \frac{1}{\pi^2} \int_{0^+}^\infty
\frac{p^2 \cos r p}{ e^{bp^2} - 1}dp \right |
\end{aligned}
\label{lowtb}
\end{equation}
where the magnon number in ground state is $N_0 =N  [ 1 - \left ( T/T_b \right )^{3/2} ]$.
Figure \ref{fig2} shows the concurrence as a function of temperature
in a system with $T_b/T_c = 0.08$ (blue line) and 0.1 (yellow line),
respectively. The concurrence keeps decreasing as the temperature
increases, this is a result of the interplay between thermal and
condensed magnons around the critical temperature and it is quite
different from magnon gas with unconserved numbers as shown in
Fig. \ref{fig1}. For $T<T_b$, the condensed magnon density $N_0/V$
decreases with the increase of temperature and results in a decreasing
temperature dependence of concurrence as indicated in Eq. (\ref{lowtb})
while the excited magnon density contributes negatively. The two types
of magnons compete with each other in determining the two-spin entanglement.
The decreasing trend illustrated in Fig. \ref{fig2} suggests that the
condensed magnons dominate the contribution to concurrence.
For $T>T_b$, only thermal magnons exist in the system. In this regime,
$g_\nu (z) \approx z$, the solution of Eq. (\ref{consistent}) gives
$z^{-1} \approx T^{3/2}$ and $n_q=1/(z^{-1} e^{bp^2} - 1) \approx T^{-1/2}$,
then the magnon density $n_q$  will decrease as temperature increases and
the bi-spin entanglement decreases accordingly.

Different from the case of thermal magnons, the entanglement does not
disappear when $T \rightarrow 0$ for condensed magnons due to the
existence of a finite magnon density under the influence of external
pumping. Furthermore, the temperature dependence of $-dC_r/dT$ is shown
as dashed lines in Fig. \ref{fig2}. It takes on a maximum value at the
condensation temperature, hence the second order derivative $-dC_r^2/dT^2$
is discontinuous at $T_b$. This provides an alternative way to characterize 
phase transition between the normal phase and the condensed phase.

Interestingly, the phase transition between normal phase and condensed
phase can be also interpreted in terms of off-diagonal long-range
order (ODLRO). Figure \ref{fig3} shows the two-spin entanglement/1-RDM
($C_{ij}=\langle a_i^\dagger a_j \rangle$) as a function of the distance
between two spins. For $T < T_b$, the 1-RDM first decays and then
saturates at a finite value as $r \rightarrow \infty$. This behavior
testifies the existence of ODRLO and further suggests that all the spins
become long-range correlated in the condensed phase. This phenomenon
is very similar to the Bose-Einstein atom condensate and superfluid
\cite{Lieb2000}. For $T > T_b$, the 1-RDM approaches zero as
$r \rightarrow \infty$, hence there is no ODLRO and the spins are only
correlated in a short range in the normal phase.

It is insightful to point out that our results
should be still valid qualitatively in the higher temperature for magnetic systems
with high Curie temperature and those materials with large spin wave gap,
where the excited magnons are still much smaller than the total number of spins.
Then we can predict that the finite entanglement can survive at the higher
temperature and this may be useful for quantum information science.

{\it\color{red}Conclusions.---} In conclusion, we have investigated
the two-spin entanglement around the critical temperature of magnon BEC.
The two-spin entanglement could be approximated as the one-particle reduced
density matrix of the system. Under condensation temperature, the
entanglement even exists when the two spins are separated by an infinitely
large distance. This suggests the existence of ODLRO in the magnon
condensate, where the long-range entanglement of spins is determined by
the significant occupation of ground state. As temperature increases
above condensation temperature, the ODLRO vanishes. Around the
condensation point, the second order derivative of entanglement
with temperature takes on a minimum at the condensation temperature,
which provides a new indicator of the phase transition from normal
magnon gas to a condensed phase. 

This work was financially supported by National Natural
Science Foundation of China (Grants No. 61704071 and No. 11405093),
the Guangdong Innovative and Entrepreneurial Research Team Program (No. 2016ZT06D348),
and the Science Technology and Innovation Commission of Shenzhen Municipality
(ZDSYS20170303165926217, JCYJ20170412152620376).


\begin{thebibliography}{}
\bibitem{note02} The form of density matrix is applicable to any systems
with rotational symmetry around $z$-axis, which is not restricted
to the Hamiltonian we consider here.
\bibitem{note01} Here if $\lambda_3 > \lambda_1$, the concurrence
is zero. Equation (5) also gives zero for this case. Then we can
always use Eq. (5)  to calculate the concurrence.
\end{thebibliography}
\end{document}